\documentclass[prl,twocolumn,showpacs,a4paper,superscriptaddress]{revtex4}%
\usepackage{amssymb}
\usepackage{epsfig}
\usepackage{amsmath}
\usepackage{graphicx}
\usepackage{subfigure}
\usepackage{amsfonts}

\newcommand{\rmB}{\mathrm{B}}

\newcommand{\rmd}{\mathrm{d}}

\newcommand{\rmm}{\mathrm{m}}

\newcommand{\rmin}{\mathrm{in}}
\newcommand{\rmout}{\mathrm{out}}

\newcommand{\bfk}{\mathbf{k}}
\newcommand{\bfd}{\mathbf{d}}
\newcommand{\trp}{\mathsf{T}}

\newcommand{\rmcon}{\mathrm{con}}
\newcommand{\bfO}{\mathbf{O}}
\newcommand{\bfR}{\mathbf{R}}
\newcommand{\bfV}{\mathbf{V}}
\newcommand{\bfX}{\mathbf{X}}
\newcommand{\bfZ}{\mathbf{Z}}
\newcommand{\bfB}{\mathbf{B}}
\newcommand{\bfC}{\mathbf{C}}
\newcommand{\bfD}{\mathbf{D}}
\newcommand{\bfE}{\mathbf{E}}
\newcommand{\bfF}{\mathbf{F}}
\newcommand{\bfI}{\mathbf{I}}
\newcommand{\bfM}{\mathbf{M}}

\begin{document}
\title{Entanglement swapping with local certification: \\Application to remote micromechanical resonators}
\author{M. Abdi}
\affiliation{Department of Physics, Sharif University of Technology, Tehran, Iran}
\affiliation{School of Science and Technology, Physics Division, University of Camerino,
Camerino (MC), Italy}
\author{S. Pirandola}
\affiliation{Department of Computer Science, University of York, York, United Kingdom}
\author{P. Tombesi}
\author{D. Vitali}
\affiliation{School of Science and Technology, Physics Division, University of Camerino,
Camerino (MC), Italy}

\begin{abstract}
We propose a protocol for entanglement swapping which involves tripartite
systems. The generation of remote entanglement induced by the Bell measurement
can be easily certified by additional local measurements. We illustrate the
protocol in the case of continuous variable systems where the certification is
effective for an appropriate class of three-mode Gaussian states. We then
apply the protocol to optomechanical systems, showing how mechanical
entanglement between two remote micromechanical resonators can be generated
and certified via local optical measurements.

\end{abstract}

\maketitle

The protocol of entanglement swapping~\cite{RMP} is recognized as a
fundamental tool for quantum communication. It is one of the central
mechanisms at the basis of quantum repeaters~\cite{Repeaters}, which allow
to distribute entanglement at long distances in an efficient manner. Very
recently, the importance of entanglement swapping has been also recognized in
quantum cryptography, where this technique introduces a Hilbert space
filtering able to exclude side-channel attacks~\cite{SideChannels}.

Entanglement swapping allows to entangle two distant systems which never
interacted. The standard protocol considers two distant users, Alice and Bob,
each possessing a bipartite entangled system, $R_{1}B_{1}$ for Alice and
$R_{2}B_{2}$ for Bob. The users send their systems $B_{1}$ and $B_{2}$ to an
intermediate station (Charlie). Here an appropriate Bell measurement is made
over the $B$ systems and the outcome communicated back to Alice and Bob. As a
result, the two remote systems $R_{1}$ and $R_{2}$ become entangled.

After the completion of this procedure, an important point is the experimental
verification of the generated entanglement, which generally involves direct
measurements on the two remote systems, $R_{1}$ and $R_{2}$, followed by
classical communications. This is fine as long as systems $R_{1}$ and $R_{2}$
are easily measurable, which is not always the case though. For instance, an
important scenario is represented by the entanglement swapping of
optomechanical systems, where $R_{1}$ and $R_{2}$ are mechanical modes while
$B_{1}$ and $B_{2}$ are optical modes~\cite{Pirandola2006}. In this case,
entanglement generation is easy, since the Bell measurement is done on the
optical modes, but entanglement verification is much more difficult, since it
involves quantum-limited measurements on mechanical modes~\cite{Kippenberg2007,Genes2009,Aspelmeyer2010,Paternostro2010,DeChiara2011}, furthermore placed at distant nodes.

Whenever the verification of entanglement involves difficult quantum
measurements, an experimentalist may therefore consider an alternative
approach which is based on the local certification protocol proposed in this
Letter. The key idea is to extend the initial systems of Alice and Bob from
bipartite to tripartite. We add a certification system ($C$) to each site, so
that we have tripartite system $R_{1}B_{1}C_{1}$ for Alice and another
$R_{2}B_{2}C_{2}$ for Bob. In this new protocol, also the certification
systems $C_{1}$ and $C_{2}$ are sent to Charlie's station, where they are
locally measured. This additional measurement enables Charlie to
\textit{certify} the generation of entanglement for the two remote systems
$R_{1}$ and $R_{2}$.

In practical implementations, the certification systems are chosen to be easy
to detect (e.g., ancillary optical modes in the case of optomechanical
systems). However, in order to achieve certification, we cannot use arbitrary
tripartite states but a suitable class of states must be engineered, that we
call \textquotedblleft certifying\textquotedblright. In our derivation, where
we consider continuous variable systems, we show that these states correspond
to a generic class of three-mode Gaussian states satisfying specific conditions of purity.

As an application of our protocol, we consider the scenario where Alice and
Bob possess two identical optomechanical systems, each formed by two output modes
of an optical cavity coupled to a micro-mechanical resonator. One traveling
optical output mode from each site is mixed and employed for the Bell
measurement generating entanglement between the two remote micro-mechanical
resonators, while the ancillary optical output mode from each site is used for
entanglement certification.
The protocol proposed here is easier to implement than a standard entanglement swapping protocol based on bipartite optomechanical states at Alice and Bob sites. In fact, the preparation of the two initial certifying states is not more difficult than preparing bipartite optomechanical entanglement, because adding a second driven cavity mode is a standard option in current cavity optomechanics~\cite{Aspelmeyer2010}. Instead, the experimental verification stage is much simpler because it involves only \emph{local optical} measurements at Charlie's site, avoiding any nonlocal measurement on the mechanical resonators.
We show that using state-of-the-art apparatuses, it
is possible to generate robust entanglement between two remote mesoscopic
systems, which can be easily certified by local optical measurements. Such a
scheme could be exploited for fundamental tests of quantum mechanics at the
macroscopic level \cite{Marshall2003,Romero-Isart2011}.

The Letter is organized as follows. We start with the definition of tripartite
certifying states. Then, we describe the entanglement swapping protocol with
local certification. Finally, we apply the protocol to the case of
optomechanical systems.

\textit{Tripartite certifying state}s.~~Consider a bosonic system which is
composed of three modes, labelled as $R$ (for remote mode), $B$ (for Bell
mode), and $C$ (for certification mode). Each mode is described by quadrature
operators $\hat{x}_{k}$ and $\hat{p}_{k}$, with commutation relations
$[\hat{x}_{k},\hat{p}_{k^{\prime}}]=i\delta_{kk^{\prime}}$ and $[\hat{x}%
_{k},\hat{x}_{k^{\prime}}]=[\hat{p}_{k},\hat{p}_{k^{\prime}}]=0$. We assume
that the system is prepared in a zero-mean Gaussian state. This means that its
characteristic function is Gaussian $\Phi(\mathbf{k})=\exp(-\mathbf{k}%
^{\mathsf{T}}\mathbf{Vk}/2)$, where $\mathbf{k}\in\mathbb{R}^{6}$ and
$\mathbf{V}$ is the covariance matrix (CM). We can always write this CM in the
form%
\begin{equation}
\mathbf{V}=\left[
\begin{array}
[c]{ccc}%
\mathbf{R} & \mathbf{D} & \mathbf{F}\\
\mathbf{D}^{\mathsf{T}} & \mathbf{B} & \mathbf{E}\\
\mathbf{F}^{\mathsf{T}} & \mathbf{E}^{\mathsf{T}} & \mathbf{C}%
\end{array}
\right]  ~,\label{CM}%
\end{equation}
where the blocks $\mathbf{R,B,C,D,E,F}$ are $2\times2$ submatrices. By
assuming non-singular blocks, this CM can always be simplified to the
standard form of Ref.~\cite{Wang2003} via local unitary operations. We implicitly assume this standard form, where $\mathbf{R}=r\mathbf{I}$,
$\mathbf{B}=b\mathbf{I}$, $\mathbf{C}=c\mathbf{I}$ ($\mathbf{I}$ is the
identity matrix) and $\mathbf{D}$, $\mathbf{E}$ are diagonal \cite{SupplMat}.
From the tripartite CM, we can extract the two submatrices
\begin{equation}
\mathbf{V}_{\mathsf{RB}}=\left[
\begin{array}
[c]{cc}%
\mathbf{R} & \mathbf{D}\\
\mathbf{D}^{\mathsf{T}} & \mathbf{B}%
\end{array}
\right]  ,~\mathbf{V}_{\mathsf{BC}}=\left[
\begin{array}
[c]{cc}%
\mathbf{B} & \mathbf{E}\\
\mathbf{E}^{\mathsf{T}} & \mathbf{C}%
\end{array}
\right]  ~,
\end{equation}
where $\mathbf{V}_{\mathsf{RB}}$ describes the remote-Bell modes, and
$\mathbf{V}_{\mathsf{BC}}$ the Bell-certification modes. Denoting with $\mu_{_{\mathsf{RB}}}=\left[4\sqrt{\det\mathbf{V}_{\mathsf{RB}}}\right]^{-1}$, $\mu_{_{\mathsf{BC}}}=\left[4\sqrt{\det\mathbf{V}_{\mathsf{BC}}}\right]^{-1}$, and $\mu_{_{\mathsf{B}}}=\left[2\sqrt{\det\mathbf{B}}\right]^{-1}$ the purities of the $RB$, $BC$, and $B$ subsystems respectively, we say that the
3-mode Gaussian state is \emph{certifying} when
\begin{equation}\label{eq:cert}
    \mu_{_{\mathsf{RB}}}>\mu_{_{\mathsf{BC}}} > \mu_{_{\mathsf{B}}}.
\end{equation}
Notice that the certifying condition implies that both $RB$ and $BC$ subsystems are entangled. In fact, it is $\mu_{_{\mathsf{B}}} \geq \mu_{_{\mathsf{B}}} \mu_{_{\mathsf{C}}}/\sqrt{\mu_{_{\mathsf{B}}}^2+\mu_{_{\mathsf{C}}}^2-\mu_{_{\mathsf{B}}}^2 \mu_{_{\mathsf{C}}}^2}$  whenever $0\leq \mu_{_{\mathsf{B}}},\mu_{_{\mathsf{C}}} \leq 1$. Therefore, for a certifying state one has $\mu_{_{\mathsf{BC}}} > \mu_{_{\mathsf{B}}} \mu_{_{\mathsf{C}}}/\sqrt{\mu_{_{\mathsf{B}}}^2+\mu_{_{\mathsf{C}}}^2-\mu_{_{\mathsf{B}}}^2 \mu_{_{\mathsf{C}}}^2}$ which, using the result of Refs.~\cite{Adesso2004a,Adesso2004b}, implies that the $BC$ subsystem is entangled. The same argument applies to the $RB$ subsystem.

\begin{figure}[t]
\centering
\includegraphics[width=0.49\textwidth]{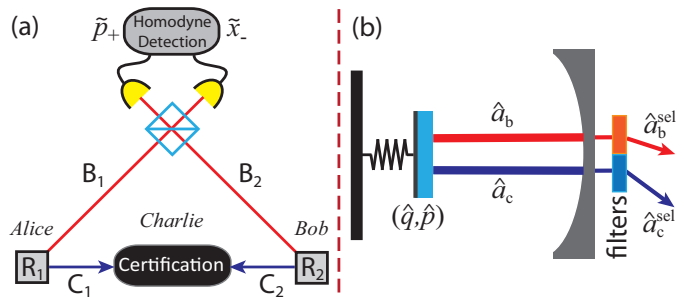} 
\caption{(Color online)
(a)~\textit{Entanglement swapping with local certification}. Alice and
Bob possess two tripartite systems, $R_{1}B_{1}C_{1}$ and $R_{2}B_{2}C_{2}$,
prepared in a certifying state. Alice and Bob keep their remote
modes $R_{1}$ and $R_{2}$, while sending the other modes to Charlie. Bell
modes $B_{1}$ and $B_{2}$ are subject to a Bell measurement with outcomes
\{$\tilde{x}_{-},\tilde{p}_{+}$\} stored by Charlie. Certification modes $C_{1}$ and $C_{2}$
are then subject to local measurements. The detection of $C_{1}C_{2}$ entanglement certifies that potential entanglement is also present
between between $R_{1}$ and $R_{2}$.
This remote entanglement is then established by the communication of \{$\tilde{x}_{-}%
,\tilde{p}_{+}$\} to Alice and Bob. (b)~\textit{Optomechanical system mounted
on each site}. The remote mode is the mechanical resonator (quadratures
$\hat{q}$ and $\hat{p}$) while the Bell and certification modes are the two
output optical modes with annihilation operators $\hat{a}_{b}^{\text{sel}}$
and $\hat{a}_{c}^{\text{sel}}$, respectively. These modes derive from
corresponding intracavity modes $\hat{a}_{b}$ and $\hat{a}_{c}$ coupled with
the mechanical resonator.}%
\label{scheme}%
\end{figure}

\textit{Entanglement swapping with certification.}~~Consider two distant
users, Alice and Bob. Each user has a system composed of three bosonic modes:
these are $R_{1}B_{1}C_{1}$ for Alice and $R_{2}B_{2}C_{2}$\ for Bob (see
Fig.~\ref{scheme}a). For simplicity we consider here the case where the two systems are prepared in the \emph{same
certifying} state with CM $\mathbf{V}_{\text{R}_{1}\text{B}_{1}\text{C}_{1}%
}=\mathbf{V}_{\text{R}_{2}\text{B}_{2}\text{C}_{2}}=\mathbf{V}$, which can be
expressed in the blockform of Eq.~(\ref{CM}).

In the first step of the protocol, Alice and Bob retain their remote modes
while sending Bell and certification modes to Charlie. In other words, Alice
keeps mode $R_{1}$, Bob keeps mode $R_{2}$, while Charlie gets modes $B_{1}$,
$B_{2}$, $C_{1}$ and $C_{2}$ (see Fig.~\ref{scheme}a).

In the second step, Charlie performs a CV Bell measurement on the Bell modes
$B_{1}$ and $B_{2}$ by using a balanced beam-splitter and two homodyne
detectors. This corresponds to measuring the two combinations of quadratures
$\hat{x}_{-}\equiv\hat{x}_{\mathrm{B}_{2}}-\hat{x}_{\mathrm{B}_{1}}$ and
$\hat{p}_{+}\equiv\hat{p}_{\mathrm{B}_{2}}+\hat{p}_{\mathrm{B}_{1}}$. The
outcomes of the measurement \{$\tilde{x}_{-},\tilde{p}_{+}$\} are classically stored by Charlie.

Note that, after the Bell measurement, the conditional state of the remaining
modes $R_{1},R_{2},C_{1}$\ and $C_{2}$\ is Gaussian with CM of the
form~\cite{SupplMat}%
\begin{equation}
\mathbf{V}_{\mathrm{out}}=\left[
\begin{array}
[c]{cc}%
\mathbf{V}_{\mathsf{R}_{1}\mathsf{R}_{2}} & \mathbf{X}\\
\mathbf{X}^{\text{T}} & \mathbf{V}_{\mathsf{C}_{1}\mathsf{C}_{2}}%
\end{array}
\right]  ~, \label{outcm}%
\end{equation}
where $\mathbf{V}_{\mathsf{R}_{1}\mathsf{R}_{2}}$ describes Alice-Bob remote
modes, $\mathbf{V}_{\mathsf{C}_{1}\mathsf{C}_{2}}$ describes the certification
modes in Charlie's hands, and $\mathbf{X}$ accounts for cross-correlations.
From the blocks, $\mathbf{V}_{\mathsf{R}_{1}\mathsf{R}_{2}}$ and
$\mathbf{V}_{\mathsf{C}_{1}\mathsf{C}_{2}}$, one computes the
partially-transposed matrices, $\mathbf{V}_{\mathsf{R}_{1}\mathsf{R}_{2}%
}^{\text{PT}}$\ and $\mathbf{V}_{\mathsf{C}_{1}\mathsf{C}_{2}}^{\text{PT}}$,
from which one can derive the minimum symplectic eigenvalues~\cite{Pirandola2006,SupplMat}
\begin{equation}
\eta_{\mathsf{R}_{1}\mathsf{R}_{2}}^{-}=\frac{\mu_{_{\mathsf{B}}}}{2\mu_{_{\mathsf{RB}}}},\;\;\;\;\;\;
\eta_{\mathsf{C}_{1}\mathsf{C}_{2}}^{-}=\frac{\mu_{_{\mathsf{B}}}}{2\mu_{_{\mathsf{BC}}}}. \label{sympl-relat}%
\end{equation}
The certifying condition of Eq.~(\ref{eq:cert}) and the definition of logarithmic negativity $E_{N}^{\mathsf{S}_{1}\mathsf{S}_{2}}=\max\{0,-\ln2\eta^{-}_{\mathsf{S}_{1}\mathsf{S}_{2}}\}$~\cite{Symplectic,Eisert2001} imply that
\begin{equation}\label{eq:fund}
E_{N}^{\mathsf{R}_{1}\mathsf{R}_{2}} > E_{N}^{\mathsf{C}_{1}\mathsf{C}_{2}} > 0.
\end{equation}
In other words, performing the entanglement swapping protocol with a certifying state allows to entangle both the $C_1C_2$ and the $R_1R_2$ bipartite subsystems, with the latter one more entangled than the former one. Therefore the detection by Charlie of entanglement in the certification subsystem $C_1C_2$ guarantees the generation of remote entanglement, i.e., between the remote modes, $R_{1}$ and $R_{2}$, in Alice's
and Bob's stations. This is the key-idea of the protocol. Charlie can locally
test the presence of entanglement between the certification modes. If this is
detected, then Charlie is sure that potential entanglement is also present
between Alice and Bob. This remote entanglement will be definitely established
once that the Bell-measurement results \{$\tilde{x}_{-},\tilde{p}_{+}$\} are communicated to
Alice and Bob.

Therefore in the third step of the protocol, soon after the Bell measurement, Charlie detects the certification modes $C_{1}$ and $C_{2}$, by measuring the quadratures or other entanglement
witness operators. After the successful detection of $C_1C_2$ entanglement, Charlie communicates the Bell-measurement outcomes
\{$\tilde{x}_{-},\tilde{p}_{+}$\} to Alice and Bob. As a result, the two users now share two
entangled modes, $R_{1}$ and $R_{2}$, which are described by a Gaussian state
with CM\ $\mathbf{V}_{\mathsf{R}_{1}\mathsf{R}_{2}}$ and non-zero first
moments $\mathbf{d}_{\mathsf{R}_{1}}$ and $\mathbf{d}_{\mathsf{R}_{2}}$. These
moments depend on the pair \{$\tilde{x}_{-},\tilde{p}_{+}$\} and are known to Alice and Bob.
Thus, the two users may decide to delete $\mathbf{d}_{\mathsf{R}_{1}}$ and
$\mathbf{d}_{\mathsf{R}_{2}}$ by applying displacements~\cite{SupplMat}. More
simply, they store the values of the first moments and account for them in the
classical post-processing of data. The previous steps can be repeated many
times, with Alice and Bob collecting a set of entangled modes \{$R_{1}^{(1)}%
$,$R_{2}^{(1)}$\}, \{$R_{1}^{(2)}$,$R_{2}^{(2)}$\}, $\cdots$, where each pair
\{$R_{1}^{(i)}$,$R_{2}^{(i)}$\} has the same CM\ $\mathbf{V}_{\mathsf{R}%
_{1}\mathsf{R}_{2}}$ but different first moments \{$\mathbf{d}_{\mathsf{R}%
_{1}}^{(i)}$,$\mathbf{d}_{\mathsf{R}_{2}}^{(i)}$\}.

\textit{Application to optomechanical systems}.~~As a first relevant
application of the proposed protocol, we consider the case where the initial
tripartite systems are two identical cavity optomechanical systems
\cite{Kippenberg2007,Genes2009,Aspelmeyer2010,Borkje2011} [see Fig.~(\ref{scheme}b)]. Each system is
formed by a micro-mechanical resonator playing the role of the remote mode,
coupled to two independently driven optical cavity modes with different frequencies~\cite{Paternostro2007}, whose output field
are sent to Charlie's site (these corresponds to the Bell and certification
modes). In such an example, the proposed protocol allows to generate an
entangled state between two remote mechanical resonators which can be used for
tests of quantum mechanics on massive objects at the nanogram level, and whose
entanglement can be certified using only standard optical homodyne
measurements at Charlie's site. In Fig.~(\ref{scheme}c) we have schematically
described the optomechanical system as a bi-chromatically driven Fabry-Perot
cavity with a movable micromirror, but the present treatment applies to a
generic system in which the optomechanical coupling is linear in the resonator
position (e.g., see Refs.~\cite{Kippenberg2007,Genes2009,Aspelmeyer2010}).

The Hamiltonian of the optomechanical system is
\begin{equation}
H=H_{\mathrm{O}}+H_{\mathrm{M}}+H_{\mathrm{OM}}+H_{\mathrm{IN}}, \label{hamil}%
\end{equation}
where $H_{\mathrm{O}}=\sum\hbar\omega_{k}\hat{a}_{k}^{\dagger}\hat{a}_{k}$
with $k=\mathrm{b},\mathrm{c}$ is the Hamiltonian of the two driven cavity
fields, and $H_{\mathrm{M}}=\hbar\omega_{\mathrm{m}}(\hat{p}^{2}+\hat{q}%
^{2})/2$ is the Hamiltonian of the mechanical resonator with mass $m$,
resonance frequency $\omega_{\mathrm{m}}$, and dimensionless quadratures
$\hat{p}$ and $\hat{q}$ with $[\hat{q},\hat{p}]=i$. The third term represents
the optomechanical interaction $H_{\mathrm{OM}}=-\hbar\hat{q}\sum G_{0k}%
\hat{a}_{k}^{\dagger}\hat{a}_{k}$, with coupling constants $G_{0k}=(\omega
_{k}/L)\sqrt{\hbar/m\omega_{\mathrm{m}}}$ (for a Fabry-Perot cavity of length
$L$ with a movable micromirror). The last term describes cavity driving,
$H_{\mathrm{IN}}=i\hbar\sum\mathcal{E}_{k}\hat{a}_{k}^{\dagger}\exp
(-i\omega_{\mathrm{L},k}t)+c.c.$, with driving rate $|\mathcal{E}_{k}%
|=\sqrt{2\kappa_{k}P_{k}/\hbar\omega_{\mathrm{L},k}}$, where $\omega
_{\mathrm{L},k}$ is the frequency of the input laser, $P_{k}$ is its power,
and $\kappa_{k}$ is the cavity mode loss rate through its input port.
The above Hamiltonian is valid when the scattering of photons of the driven modes into other cavity modes and also between the two chosen modes is negligible: this is valid when $\omega_{\rmm}$ is much smaller then the free spectral range of the cavity \cite{Genes2009}.

The full dynamics of the cavity optomechanical system involves
fluctuation-dissipation processes, which can be taken into account by adopting
quantum Langevin equations (QLEs) \cite{Giovannetti2001}. If both cavity modes
are intensely driven and the system is in its stability region, the latter
reaches a semiclassical steady state with large intracavity amplitudes for
both cavity modes, $\alpha_{\mathrm{s},k}$, and a shifted equilibrium position
of the resonator, $q_{\mathrm{s}}$. The fluctuations of the various operators
around such a steady state obey a quantum dynamics well described by a
linearization of the QLE. This corresponds to writing an operator
$\hat{\mathcal{O}}$ as the sum of its semiclassical value plus quantum
fluctuations $\hat{\mathcal{O}}=\mathcal{O}_{\mathrm{s}}+\delta\hat
{\mathcal{O}}$~\cite{Genes2009}. The linearized QLEs have a compact form in
terms of the quadrature fluctuations of the mechanical mode \{$\delta\hat
{q},\delta\hat{p}$\} and the two cavity modes, $\delta\hat{x}_{k}$ and
$\delta\hat{y}_{k}$, defined by the relation $\delta\hat{a}_{k}\equiv
(\delta\hat{x}_{k}+i\delta\hat{y}_{k})/\sqrt{2}$. In fact, the QLEs can be
written as $\dot{\hat{\mathbf{u}}}=\mathcal{A}\hat{\mathbf{u}}+\hat
{\mathbf{n}}$, where $\hat{\mathbf{u}}=(\delta\hat{q},\delta\hat{p},\delta
\hat{x}_{\mathrm{b}},\delta\hat{y}_{\mathrm{b}},\delta\hat{x}_{\mathrm{c}%
},\delta\hat{y}_{\mathrm{c}})^{\mathsf{T}}$. In this formula, $\hat
{\mathbf{n}}=(0,\hat{\xi},\sqrt{2\kappa}\hat{x}_{\mathrm{b}}^{\mathrm{in}%
},\sqrt{2\kappa}\hat{y}_{\mathrm{b}}^{\mathrm{in}},\sqrt{2\kappa},\hat
{x}_{\mathrm{c}}^{\mathrm{in}},\sqrt{2\kappa}\hat{y}_{\mathrm{c}}%
^{\mathrm{in}})^{\mathsf{T}}$ is a noise vector, where $\hat{\xi}$ is the
thermal noise affecting the resonator, and \{$\hat{x}_{\mathrm{k}%
}^{\mathrm{in}},\hat{y}_{\mathrm{k}}^{\mathrm{in}}$\} are optical input
noises. Finally, $\mathcal{A}$ is the drift matrix of the system quantum
fluctuations, which is given by
\begin{equation}
\mathcal{A}=\left[
\begin{array}
[c]{cccccc}%
0 & \omega_{\mathrm{m}} & 0 & 0 & 0 & 0\\
-\omega_{\mathrm{m}} & \gamma_{\mathrm{m}} & G_{\mathrm{b}} & 0 &
G_{\mathrm{c}} & 0\\
0 & 0 & -\kappa_{\mathrm{b}} & \Delta_{\mathrm{b}} & 0 & 0\\
G_{\mathrm{b}} & 0 & -\Delta_{\mathrm{b}} & -\kappa_{\mathrm{b}} & 0 & 0\\
0 & 0 & 0 & 0 & -\kappa_{\mathrm{c}} & \Delta_{\mathrm{c}}\\
G_{\mathrm{c}} & 0 & 0 & 0 & -\Delta_{\mathrm{c}} & -\kappa_{\mathrm{c}}%
\end{array}
\right]  ,
\end{equation}
where $\gamma_{m}$ is the mechanical damping rate, $G_{k}=\sqrt{2}G_{0k}%
\alpha_{\mathrm{s},k}$ are the effective optomechanical coupling between the
fluctuations, and $\Delta_{k}=\omega_{k}-\omega_{\mathrm{L},k}-G_{0k}%
q_{\mathrm{s}}$ are the effective detunings of the two cavity modes.

We are however interested in the traveling output fields associated with the
cavity modes. These fields are sent to Charlie for the Bell and subsequent
certifying measurements. Considering the cavity output is advantageous also
for achieving larger and more robust entanglement: in fact, by properly
choosing the central frequency and the bandwidth of the output modes, one can
\emph{optimally filter} the frequencies carrying the proper optomechanical
quantum correlations \cite{Genes2009,Genes2008}. The selected output modes are
defined by the following bosonic annihilation operators $\hat{a}%
_{k}^{\mathrm{sel}}(t)=\int_{t_{0}}^{t}h_{k}(t-s)\hat{a}_{k}^{\mathrm{out}%
}(s)\mathrm{d}s$, where $\hat{a}_{k}^{\mathrm{out}}(t)=\sqrt{2\kappa}%
\delta\hat{a}_{k}(t)-\hat{a}_{k}^{\mathrm{in}}(t)$ is the standard
input-output relation for the optical field, and $h_{k}(t)$ is a causal filter
function defining the output modes~\cite{Genes2008}. In fact, $\hat{a}%
_{k}^{\mathrm{sel}}$ is a standard photon annihilation operator, implying the
normalization condition $\int|h_{k}(t)|^{2}\mathrm{d}t=1$. A simple choice is
taking $h_{k}(t)=\sqrt{2/\tau_k}\Theta(t)\exp[-(1/\tau_k+i\Omega_{k})t]$ where
$\Theta(t)$ is the Heaviside step function, $1/\tau_k$ is the bandwidth of the
filter, and $\Omega_{k}$ is the central frequency
(measured with respect to the frequency of the corresponding driving field).

We are interested in the stationary state of the quadrature fluctuations of
the mechanical resonator and the two selected output modes. This is a
zero-mean Gaussian state since the system is driven by Gaussian noises and the
dynamics is linearized around a semiclassical point. This tripartite Gaussian
state, for both Alice's and Bob's sites, represents the input state of our
entanglement swapping protocol. It is fully characterized by the CM
$\mathbf{V}_{ij}^{\mathrm{out}}=\langle\hat{u}_{i}^{\mathrm{out}}(\infty
)\hat{u}_{j}^{\mathrm{out}}(\infty)+\hat{u}_{j}^{\mathrm{out}}(\infty)\hat
{u}_{i}^{\mathrm{out}}(\infty)\rangle/2$, where $\hat{\mathbf{u}%
}^{\mathrm{out}}=(\delta\hat{q},\delta\hat{p},\hat{x}_{\mathrm{b}%
}^{\mathrm{sel}},\hat{y}_{\mathrm{b}}^{\mathrm{sel}},\hat{x}_{\mathrm{c}%
}^{\mathrm{sel}},\hat{y}_{\mathrm{c}}^{\mathrm{sel}})^{\mathsf{T}}$, whose
form can be explicitly evaluated using the formalism of
Refs.~\cite{Genes2009,Genes2008}.

\begin{figure}[ptbh]
\centering \vspace{+0.3cm} \includegraphics[width=3.2in]{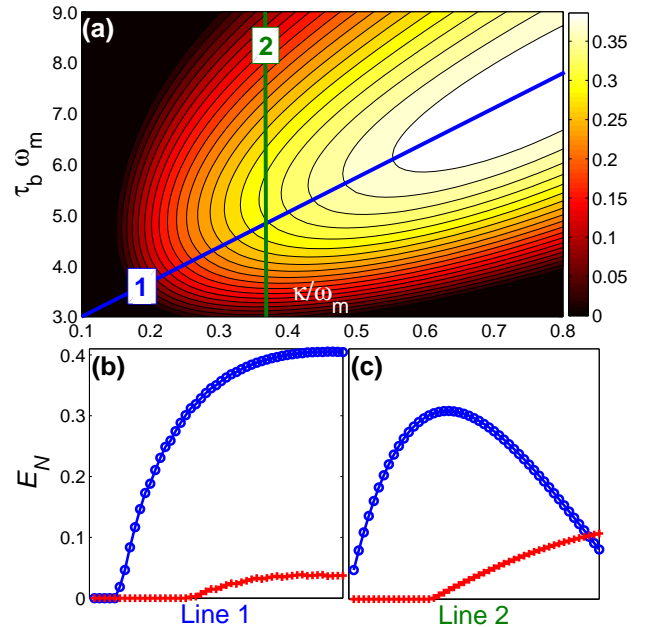}
\vspace{-0.3cm} \caption{(Color online) (a)~Contour plot of the logarithmic negativity
$E_{N}$ of the remote mechanical resonators with respect to the normalized
inverse filter bandwidth $\tau_b\omega_{m}$, and the normalized cavity decay rate
$\kappa/\omega_{m}$. (b) and~(c) show $E_{N}$ of the mechanical
resonators (blue circle) and of the certification modes (red plus) along line $1$
and line $2$, respectively. Parameters are: Cavity
length $L=1$~mm, mechanical resonator with frequency $\omega_{\mathrm{m}%
}/2\pi=10$~MHz, quality factor $Q_{\mathrm{m}}=10^{6}$, and mass $m=10$~ng,
giving coupling constants $G_{0k}\approx1$~KHz. Optical modes have detuning
values $\Delta_{\mathrm{c}}=-\Delta_{\mathrm{b}}=\omega_{\mathrm{m}}$, while the output modes have
central frequencies $\Omega_{\mathrm{b}}=-\omega_{\mathrm{m}}$ and
$\Omega_{\mathrm{c}}=\omega_{\mathrm{m}}$, the relation between the two inverse bandwidths being $\tau_{c}=\tau_{b}/5$.  Input laser powers are
$P_{\mathrm{b}}=4.5~\mathrm{mW}$ and $P_{\mathrm{c}}=5.0~\mathrm{mW}$, with
$\lambda_{\mathrm{b}}\approx\lambda_{\mathrm{c}}=810$~nm. Reservoir
temperature is $T=0.1$~K.}%
\label{logneg}%
\end{figure}

The system parameters must be chosen in such a way that the tripartite system formed by the mechanical resonator and the two cavity output fields is in a stationary state satisfying the certifying condition of Eq.~(\ref{eq:cert}) which, as noticed above, implies the presence of nonzero entanglement between the mechanical resonator and the output mode employed for the Bell measurement, and also between the two optical output fields.  The certifying condition is satisfied if we appropriately choose the
detunings and filter the appropriate output modes. In fact, the optomechanical
entanglement is largest when we drive the cavity Bell mode with a blue-detuned
laser ($\Delta_{\mathrm{b}}<0$) and the cavity certifying mode by a
red-detuned laser ($\Delta_{\mathrm{c}}>0$). The behavior of the system in an experimentally feasible parameter region is shown in Fig.~\ref{logneg}(a), where the resulting log-negativity $E_{N}$ between the two remote mechanical resonators is plotted as a function of the normalized
inverse bandwidth of the Bell output mode $\tau_b\omega_{m}$, and of the cavity mode
bandwidth $\kappa/\omega_{m}$ (equal for the two modes). Figs.~\ref{logneg}(b),(c) show $E_{N}^{\mathsf{R}_{1}\mathsf{R}_{2}}$ of the remote mechanical
resonators (blue circles) and $E_{N}^{\mathsf{C}_{1}\mathsf{C}_{2}}$ of the certification modes (red plus) along lines $1$
and $2$ of Fig.~\ref{logneg}(a), respectively. In Figs.~\ref{logneg}(b),(c), the parameter regions where the condition $E_{N}^{\mathsf{R}_{1}\mathsf{R}_{2}} > E_{N}^{\mathsf{C}_{1}\mathsf{C}_{2}} > 0$ of Eq.~(\ref{eq:fund}) is satisfied correspond to the situation where the initial optomechanical tripartite state satisfies the certifying condition.

In conclusion, we have proposed an extension of entanglement swapping protocol
where the generation of entanglement between two remote systems can be
certified by local measurements performed soon after the Bell measurement. We
have then shown that this protocol can be applied for entangling two distant
mechanical resonators, which are part of two optomechanical systems. The
resulting mechanical entanglement can be successfully certified by applying
local optical measurements.

\begin{appendix}

\section{Supplementary Material}
%
%
\subsection{Conditional correlation matrix}
The system is initially composed of two distant, independent, tripartite continuous variable (CV) systems, $R_{1}B_{1}C_{1}$ in
Alice's hands, and $R_{2}B_{2}C_{2}$ in Bob's hands.
As a consequence, the total initial state is given by tensor product of the states prepared by Alice and Bob, $\rho_{\rmin}=\rho_{1}\otimes\rho_{2}$, which in
the Wigner function formalism, can be expressed as the product of the Wigner functions of the two subsystems
\begin{equation}
W_{\rmin} =W_{1}(\alpha_{1},\beta_{1},\gamma_{1})W_{2}(\alpha_{2},\beta_{2},\gamma_{2}) ~,
\end{equation}
where $\alpha_{k}$, $\beta_{k}$, and $\gamma_{k}$, with $k=1,2$, are the complex amplitudes related to the phase space variables, e.g. we have $\alpha_{k} \equiv (x_{k}+ip_{k})/\sqrt{2}$.

Then, both Alice and Bob send two of their modes, the Bell modes $B_{k}$, and the certifying modes $C_{k}$, to Charlie, while holding the remote modes $R_{k}$ in their hands. At the next step, Charlie performs the CV version of the Bell measurement on the two Bell modes $B_1$ and $B_2$:
he mixes $B_{1}$ and $B_2$ with a balanced beam splitter giving the following output modes
\begin{equation}
\hat{x}_{\pm}=\frac{\hat{x}_{\rmB_{2}} \pm \hat{x}_{\rmB_{1}}}{\sqrt{2}} ~,~~~
\hat{p}_{\pm}=\frac{\hat{p}_{\rmB_{2}} \pm \hat{p}_{\rmB_{1}}}{\sqrt{2}} ~,
\end{equation}
which are then both subjected to homodyne detection so that the quadratures $\hat{x}_{-}$ and $\hat{p}_{+}$ are measured, with outcomes $\tilde{x}_{-}$ and $\tilde{p}_{+}$ and the associated probability $P(\tilde{x}_{-},\tilde{p}_{+})$.
Such a measurement collapses the system from a hexapartite to a fully correlated quadripartite system in a state which is conditioned to the measurement outcomes. In the Wigner function formalism this is expressed by an integration over a complex amplitude mixed by the beam splitter, weighted by the measurement results probability, i.e.,
\begin{equation}\label{wigcon}
W_{\rmcon}=\frac{1}{P(\tilde{\beta})}\int\rmd^2\beta W_{1}(\alpha_{1},\gamma_{1},\beta)W_{2}(\alpha_{2},\gamma_{2},\beta^*-\tilde{\beta}^*) ~,
\end{equation}
where $\tilde{\beta} \equiv i\tilde{p}_{+} -\tilde{x}_{-}$ is the complex variable representing the measurement outcomes, and $\beta$ is the unobserved variable which is integrated out. The variable $\beta$ is related to the Bell modes variables mixed by the beam splitter, $\beta_{\pm} \equiv (\beta_{2} \pm \beta_{1})/\sqrt{2}$, with $\beta_{\pm}=(x_{\pm}+ip_{\pm})/\sqrt{2}$, by the following relation
$\beta = [x_{+} +i\tilde{p}_{+} -(\tilde{x}_{-} +ip_{-})]/2$.

An equivalent description of the conditional state is provided by its symmetrically ordered characteristic function, which is the complex Fourier transform of Wigner function $\Phi(\lambda)\equiv\mathcal{F}[W(\alpha)]$ (see e.g. Ref.~\cite{Scully1996}). Using Eq.~(\ref{wigcon}) one gets
\begin{eqnarray}\label{eq:ch-fun}
&& \Phi_{\rmcon}(\lambda_{1},\lambda_{2},\mu_{1},\mu_{2}) \\
&& =\frac{1}{\pi^{2}P(\tilde{\beta})}\int \rmd^{2}\nu_{2}  \Phi_{1}(\lambda_{1},\mu_{1},\nu_{2}^{*}) \Phi_{2}(\lambda_{2},\mu_{2},\nu_{2}) ~, \nonumber
\end{eqnarray}
where $\lambda_{1}$, $\lambda_{2}$, $\mu_{1}$, $\mu_{2}$, $\nu_{1}$, and $\nu_{2}$ are the complex variables conjugated to the Wigner function variables $\alpha_{1}$, $\alpha_{2}$, $\gamma_{1}$, $\gamma_{2}$, $\beta_{1}$, and $\beta_{2}$ respectively.

In the case considered in the Letter, the initial tripartite CV states at Alice and Bob sites are Gaussian, i.e., with symmetrically ordered characteristic function $\Phi_j(\bfk)=\exp\{-\bfk^{\trp} \bfV_j \bfk/2 +i \bfd_j^{\trp}\bfk\}$, $j=1,2$, with covariance matrix (CM)
\begin{equation}\label{incm}
\bfV_{j} =\left[\begin{array}{ccc}
    \bfR_{j} & \bfD_{j} & \bfF_{j} \\
    \bfD_{j}^{\trp} & \bfB_{j} & \bfE_{j} \\
    \bfF_{j}^{\trp} & \bfE_{j}^{\trp} & \bfC_{j}
    \end{array}\right] ~
\end{equation}
(the blocks $\bfR_{j},\bfB_{j},\bfC_{j},\bfD_{j},\bfE_{j}$, and $\bfF_{j}$ are $2\times2$ submatrices), vector of the real-valued system variables $\bfk$, and displacement vector of the state $\bfd_j$, which can be taken equal to zero (displacements do not affect entanglement).
Inserting these expressions into Eq.~(\ref{eq:ch-fun}) and performing the Gaussian integrals, one can express the effect of the Bell measurement performed by Charlie as a relation between the CM of the input and output Gaussian states. In fact, the initial state in possession of Alice and Bob is fully specified by the input $12 \times 12$ CM
\begin{equation}\label{enscm}
\bfV_{\rmin} =\left[\begin{array}{cc}
    \bfV_{1} &  \\
     & \bfV_{2}
  \end{array}\right]
\end{equation}
with $\bfV_{j}$, $j=1,2$, specified in Eq.~(\ref{incm}). After the Bell measurement, the quadripartite system $R_1R_2C_1C_2$ is in a conditional Gaussian state with $8 \times 8$ CM
\begin{equation}\label{enscm}
\bfV_{\rmout} =\left[\begin{array}{cc}
    \bfV_{\mathsf{R_{1}R_{2}}} & \bfX \\
    \bfX^{\trp} & \bfV_{\mathsf{C_{1}C_{2}}}
  \end{array}\right] ~,
\end{equation}
where the CMs $\bfV_{\mathsf{R_{1}R_{2}}}$ and $\bfV_{\mathsf{C_{1}C_{2}}}$ of the two bipartite subsystems $R_1R_2$ and $C_1C_2$, and the matrix expressing their correlations $\bfX$ are given by
\begin{eqnarray}\label{eq:vr1r2}
\bfV_{\mathsf{R_{1}R_{2}}}&=
\left[\begin{array}{cc}
		\bfR_{1} -\bfD_{1}\bfZ\bfM^{-1}\bfZ\bfD_{1}^{\trp} & \bfD_{1}\bfZ\bfM^{-1}\bfD_{2}^{\trp} \\
		\bfD_{2}\bfM^{-1}\bfZ\bfD_{1}^{\trp} & \bfR_{2} -\bfD_{2}\bfM^{-1}\bfD_{2}^{\trp}
			\end{array}\right], \\
\bfV_{\mathsf{C_{1}C_{2}}}&=
\left[\begin{array}{cc}
		\bfC_{1} -\bfE_{1}\bfZ\bfM^{-1}\bfZ\bfE_{1}^{\trp} & \bfE_{1}\bfZ\bfM^{-1}\bfE_{2}^{\trp} \\
		\bfE_{2}\bfM^{-1}\bfZ\bfE_{1}^{\trp} & \bfC_{2} -\bfE_{2}\bfM^{-1}\bfE_{2}^{\trp}
			\end{array}\right] ~, \label{eq:vc1c2}\\
\bfX&=
\left[\begin{array}{cc}
		\bfF_{1} -\bfD_{1}\bfZ\bfM^{-1}\bfZ\bfE_{1} & \bfD_{1}\bfZ\bfM^{-1}\bfE_{2} \\
		\bfE_{1}\bfZ\bfM^{-1}\bfD_{2} & \bfF_{2} -\bfE_{2}\bfM^{-1}\bfD_{2}
			\end{array}\right] ~,			
\label{outcm}
\end{eqnarray}
where we have introduced $\bfM \equiv \bfB_{1}+\bfZ\bfB_{2}\bfZ$ and $\bfZ \equiv \mathrm{diag}[1,-1]$.
The reduced state of the remote site system $R_1R_2$ possesses a nonzero displacement vector given by
\begin{align}
\bfd_{\mathsf{R_{1}}} &=-2\bfD_{1}^{\trp}\bfZ\bfM^{-1}\bfO ~, \\
\bfd_{\mathsf{R_{2}}} &=2\bfD_{2}^{\trp}\bfM^{-1}\bfO ~,
\end{align}
where $\bfO\equiv(\tilde{x}_{-},\tilde{p}_{+})^{\trp}$ is vector of the Bell measurement outcomes.

\subsection{Symplectic eigenvalues}

We can get simplified expressions by adopting a standard form for the CMs of the initial tripartite Gaussian states at Alice and Bob sites, $\bfV_{1}$ and $\bfV_{2}$. In fact, by assuming non-singular $2\times 2$ blocks, the two CMs can always be simplified to the
standard form of Ref.~\cite{Wang2003b} via local unitary operations at Alice, Bob and Charlie's sites (therefore not affecting the initial entanglement).
This standard form is given by
\begin{equation}
\bfV =\left[\begin{array}{cccccc}
    r &   & d &  & f & f' \\
      & r &   & d' & f'' & f''' \\
    d &   & b &   & e &  \\
      & d' &   & b &   & e'\\
    f & f' & e &   & c &   \\
    f'' & f''' &   & e' &   & c
  \end{array}\right] ~.
\label{stdform}
\end{equation}
This means that we can put the $2\times 2$ blocks of Eq.~(\ref{incm}) in the form $\bfR_{k}=r_{k}\bfI$, $\bfB_{k}=b_{k}\bfI$, $\bfC_{k}=c_{k}\bfI$ ($\bfI$ is the $2\times2$ identity matrix), $\bfD_{k}=\mathrm{diag}[d_{k},d'_{k}]$, $\bfE_{k}=\mathrm{diag}[e_{k},e'_{k}]$, and
\begin{align}
&\bfF_{k}=\left[\begin{array}{cc}
		f_{k} & f'_{k} \\
		f''_{k} & f'''_{k}
		\end{array}\right] ~, \nonumber
\end{align}
where $k=1,2$.
As a consequence, the CM of the reduced states of $R_1R_2$ and $C_1C_2$ of Eqs.~(\ref{eq:vr1r2}) and (\ref{eq:vc1c2}) assume the block-diagonal form
\begin{widetext}
\begin{align}
\bfV_{\mathsf{R_{1}R_{2}}} &=
\left[\begin{array}{cccc}
	  r_{1}-\frac{d_{1}^2}{b_{1}+b_{2}} &   & \frac{d_{1}d_{2}}{b_{1}+b_{2}} &   \\
	    & r_{1}-\frac{d'^{2}_{1}}{b_{1}+b_{2}} &   & -\frac{d'_{1}d'_{2}}{b_{1}+b_{2}} \\
	  \frac{d_{1}d_{2}}{b_{1}+b_{2}} &   & r_{2}-\frac{d_{2}^{2}}{b_{1}+b_{2}} &   \\
	    & -\frac{d'_{1}d'_{2}}{b_{1}+b_{2}} &   & r_{2}-\frac{d'^{2}_{2}}{b_{1}+b_{2}}
	  \end{array}\right] ~, \\
\bfV_{\mathsf{C_{1}C_{2}}} &=
\left[\begin{array}{cccc}
	  c_{1}-\frac{e_{1}^{2}}{b_{1}+b_{2}} &   & \frac{e_{1}e_{2}}{b_{1}+b_{2}} &  \\
	    & c_{1}-\frac{e'^{2}_{1}}{b_{1}+b_{2}} &   & -\frac{e'_{1}e'_{2}}{b_{1}+b_{2}} \\
	  \frac{e_{1}e_{2}}{b_{1}+b_{2}} &   & c_{2}-\frac{e_{2}^{2}}{b_{1}+b_{2}} &  \\
	    & -\frac{e'_{1}e'_{2}}{b_{1}+b_{2}} &   & c_{2}-\frac{e'^{2}_{2}}{b_{1}+b_{2}}
	  \end{array}\right] ~.
\end{align}
\end{widetext}
One then computes the
partially-transposed matrices, $\mathbf{V}_{\mathsf{R}_{1}\mathsf{R}_{2}%
}^{\text{PT}}$\ and $\mathbf{V}_{\mathsf{C}_{1}\mathsf{C}_{2}}^{\text{PT}}$,
from which one can derive the minimum symplectic eigenvalues $\eta_{\mathsf{R}%
_{1}\mathsf{R}_{2}}^{-}$ and $\eta_{\mathsf{C}_{1}\mathsf{C}_{2}}^{-}$ which allow to quantify the remote and certifying entanglement. In the general case of different tripartite states at Alice and Bob sites the explicit expressions of $\eta_{\mathsf{R}%
_{1}\mathsf{R}_{2}}^{-}$ and $\eta_{\mathsf{C}_{1}\mathsf{C}_{2}}^{-}$ are cumbersome, but they considerably simplifies in the symmetric case of identical initial resources at the two sites, $\bfV_{1}=\bfV_{2}$, considered in the Letter. In fact, one gets $\eta_{\mathsf{R_{1}R_{2}}}^{-}=b^{-1}\sqrt{\det \bfV_{\mathsf{RB}}}$ and $\eta_{\mathsf{C_{1}C_{2}}}^{-} =b^{-1}\sqrt{\det\bfV_{\mathsf{BC}}}$, where
$\bfV_{\mathsf{RB}}$ and $\bfV_{\mathsf{BC}}$ are the CMs of the reduced Gaussian input state of the bipartite subsystems $R_1B_1$ ($R_2B_2$), and $B_1C_1$ ($B_2C_2$), respectively. Using the expressions for the purities, $\mu_{_{\mathsf{RB}}}=\left[4\sqrt{\det\mathbf{V}_{\mathsf{RB}}}\right]^{-1}$, $\mu_{_{\mathsf{BC}}}=\left[4\sqrt{\det\mathbf{V}_{\mathsf{BC}}}\right]^{-1}$, and $\mu_{_{\mathsf{B}}}=\left[2\sqrt{\det\mathbf{B}}\right]^{-1}$, one arrives at Eq.~(5) of the Letter,
\begin{equation}
\eta_{\mathsf{R}_{1}\mathsf{R}_{2}}^{-}=\frac{\mu_{_{\mathsf{B}}}}{2\mu_{_{\mathsf{RB}}}},\;\;\;\;\;\;
\eta_{\mathsf{C}_{1}\mathsf{C}_{2}}^{-}=\frac{\mu_{_{\mathsf{B}}}}{2\mu_{_{\mathsf{BC}}}}. \label{sympl-relat}%
\end{equation}
This derivation of the partial transpose minimum symplectic eigenvalues practically coincides with that of Ref.~\cite{Pirandola2006b}, which refers to entanglement swapping starting from identical bipartite Gaussian states in standard form. In fact, the tripartite standard form of Ref.~\cite{Wang2003b} implies that the $BC$ and $RB$ subsystems are simultaneously put in bipartite standard form,
\begin{align}
V_{\mathsf{RB}}&= \left[
\begin{array}{cccc}
	r &   & d &   \\
	  & r &   & d' \\
	d &   & b &   \\
	  & d' &   & b
\end{array} \right] ~, \\
V_{\mathsf{BC}}&= \left[
\begin{array}{cccc}
	b &   & e &  \\
	  & b &   & e' \\
	e &   & c &   \\
	  & e' &   & c
\end{array} \right] ~.
\end{align}
Therefore one can apply the derivation of Ref.~\cite{Pirandola2006b} to both the $RB$ and $BC$ subsystems.

\end{appendix}


\begin{thebibliography}{99}                                                                                               %


\bibitem {RMP}S. L. Braunstein, and P. van Loock, Rev. Mod. Phys. \textbf{77},
513 (2005); C. Weedbrook, S. Pirandola, R. Garcia-Patron, N. J. Cerf, T. C.
Ralph, J. H. Shapiro, and S. Lloyd, arxiv:1110.3234 (2011). To appear on RMP.

\bibitem {Repeaters}H.-J. Briegel, W. D\"{u}r, J. I. Cirac, and P. Zoller,
Phys. Rev. Lett. \textbf{81}, 5932--5935 (1998).

\bibitem {SideChannels}S. L. Braunstein, and S. Pirandola, Phys. Rev. Lett. \textbf{108}, 130502 (2012).

\bibitem {Pirandola2006}S.~Pirandola, D.~Vitali, P.~Tombesi, and S.~Lloyd,
Phys.~Rev.~Lett. \textbf{97}, 150403 (2006).

\bibitem{Kippenberg2007}
T. J. Kippenberg and K. J. Vahala, Opt. Expr. \textbf{15}, 17172 (2007).

\bibitem{Genes2009}
C. Genes, A. Mari, D. Vitali, and P. Tombesi, Adv. At. Mol. Opt. Phys. \textbf{57}, 33 (2009).

\bibitem{Aspelmeyer2010}
M. Aspelmeyer, S. Gr\"{o}blacher, K. Hammerer, and N. Kiesel, J. Opt. Soc. Am. B \textbf{27}, A189 (2010).

\bibitem{Paternostro2010}M. Paternostro, G. De Chiara, and G. M. Palma, Phys. Rev. Lett. \textbf{104}, 243602 (2010).

\bibitem{DeChiara2011}G. De Chiara, M. Paternostro, and G. M. Palma, Phys. Rev. A \textbf{83}, 052324 (2011).

\bibitem{Marshall2003}W. Marshall, C. Simon, R. Penrose, and D. Bouwmeester, Phys.
Rev. Lett. \textbf{91}, 130401 (2003).

\bibitem{Romero-Isart2011} O. Romero-Isart, A. C. Pflanzer, F. Blaser, R. Kaltenbaek, N. Kiesel, M. Aspelmeyer, J. I. Cirac, Phys. Rev. Lett. \textbf{107}, 020405 (2011)

\bibitem{Wang2003}
L.~Wang, S.~Li, and H.~Zheng, Phys.~Rev.~A \textbf{67}, 062317 (2003).

\bibitem{SupplMat}See supplementary material for more details.


\bibitem{Adesso2004a}G. Adesso, A. Serafini, and F. Illuminati, Phys. Rev. Lett. \textbf{92}, 087901 (2004).

\bibitem{Adesso2004b}G. Adesso, A. Serafini, and F. Illuminati, Phys. Rev. A \textbf{70}, 022318 (2004).

\bibitem {Symplectic}Given a two-mode Gaussian state with CM $\mathbf{V}$, the
minimum symplectic eigenvalue $\eta^{-}$ of the partially-transposed CM
$\mathbf{V}^{\text{PT}}$ is an entanglement monotone, related to the
log-negativity by $E_{N}=\max\{0,-\ln2\eta^{-}\}$~\cite{Eisert2001}.

\bibitem {Eisert2001}J. Eisert, Ph.D. thesis, University of Potsdam (2001).

\bibitem{Borkje2011} K. Borkje, A. Nunnenkamp, and S. M. Girvin, Phys. Rev. Lett. \textbf{107}, 123601 (2011).

\bibitem{Paternostro2007}M. Paternostro, D. Vitali, S. Gigan, M. S. Kim, C. Brukner,
J. Eisert, and M. Aspelmeyer, Phys. Rev. Lett. \textbf{99}, 250401 (2007).

\bibitem {Giovannetti2001}V.~Giovannetti and D.~Vitali, Phys. Rev. A
\textbf{63}, 023812 (2001).

\bibitem {Genes2008}C.~Genes, A.~Mari, P.~Tombesi, and D.~Vitali, Phys.~Rev.~A
\textbf{78}, 032316 (2008).
\end{thebibliography}

\begin{thebibliography}{9}

\bibitem{Scully1996}
M.~O.~Scully and M.~S.~Zubairy, \textit{Quantum Optics} (Cambridge University Press, 1996).

\bibitem{Wang2003b}
L.~Wang, S.~Li, and H.~Zheng, Phys.~Rev.~A \textbf{67}, 062317 (2003).

\bibitem {Pirandola2006b}S.~Pirandola, D.~Vitali, P.~Tombesi, and S.~Lloyd,
Phys.~Rev.~Lett. \textbf{97}, 150403 (2006).


\end{thebibliography}
\end{document}